\documentclass[preprint]{vgtc}               

\ifpdf
  \pdfoutput=1\relax                   
  \usepackage{graphicx}                
  \DeclareGraphicsExtensions{.pdf,.png,.jpg,.jpeg} 
\else
  \usepackage{graphicx}                
  \DeclareGraphicsExtensions{.eps}     
\fi%

\graphicspath{{figures/}{pictures/}{images/}{./}} 

\usepackage{microtype}                 
\PassOptionsToPackage{warn}{textcomp}  
\usepackage{textcomp}                  
\usepackage{mathptmx}                  
\usepackage{times}                     
\usepackage{cite}                      
\usepackage{tabu}                      
\usepackage{booktabs}                  

\usepackage{xcolor}
\usepackage{multirow}
\usepackage{booktabs}
\usepackage{algorithm}
\usepackage[noend]{algpseudocode}
\usepackage{graphicx}
\usepackage{subfig}
\usepackage{enumitem}
\usepackage{lineno}

\onlineid{0}

\vgtccategory{Research}

\vgtcinsertpkg

\preprinttext{To appear in IEEE VIS 2019 Short Papers conference proceedings and IEEE Xplore.}

\title{Analyzing Time Attributes in Temporal Event Sequences}

\author{Jessica~Magallanes\thanks{e-mail: jgmagallanescastaneda1@sheffield.ac.uk}\\ %
        \parbox{1.4in}{\scriptsize \centering Department of Computer Science\\INSIGNEO Institute for \textit{in silico} Medicine\\University of Sheffield\\Sheffield, United Kingdom} %
\and Lindsey~van Gemeren\\ %
     \parbox{1.4in}{\scriptsize \centering Sheffield Teaching Hospitals \\NHS Foundation Trust\\INSIGNEO Institute for \textit{in silico} Medicine\\University of Sheffield\\Sheffield, United Kingdom}  %
\and Steven~Wood\\ %
     \parbox{1.4in}{\scriptsize \centering Sheffield Teaching Hospitals \\NHS Foundation Trust\\INSIGNEO Institute for \textit{in silico} Medicine\\University of Sheffield\\Sheffield, United Kingdom}  %
\and Maria-Cruz~Villa-Uriol\thanks{e-mail: m.villa-uriol@sheffield.ac.uk}\\ %
     \parbox{1.4in}{\scriptsize \centering Department of Computer Science\\INSIGNEO Institute for \textit{in silico} Medicine\\University of Sheffield\\Sheffield, United Kingdom}}

\teaser{
\includegraphics[width=\columnwidth]{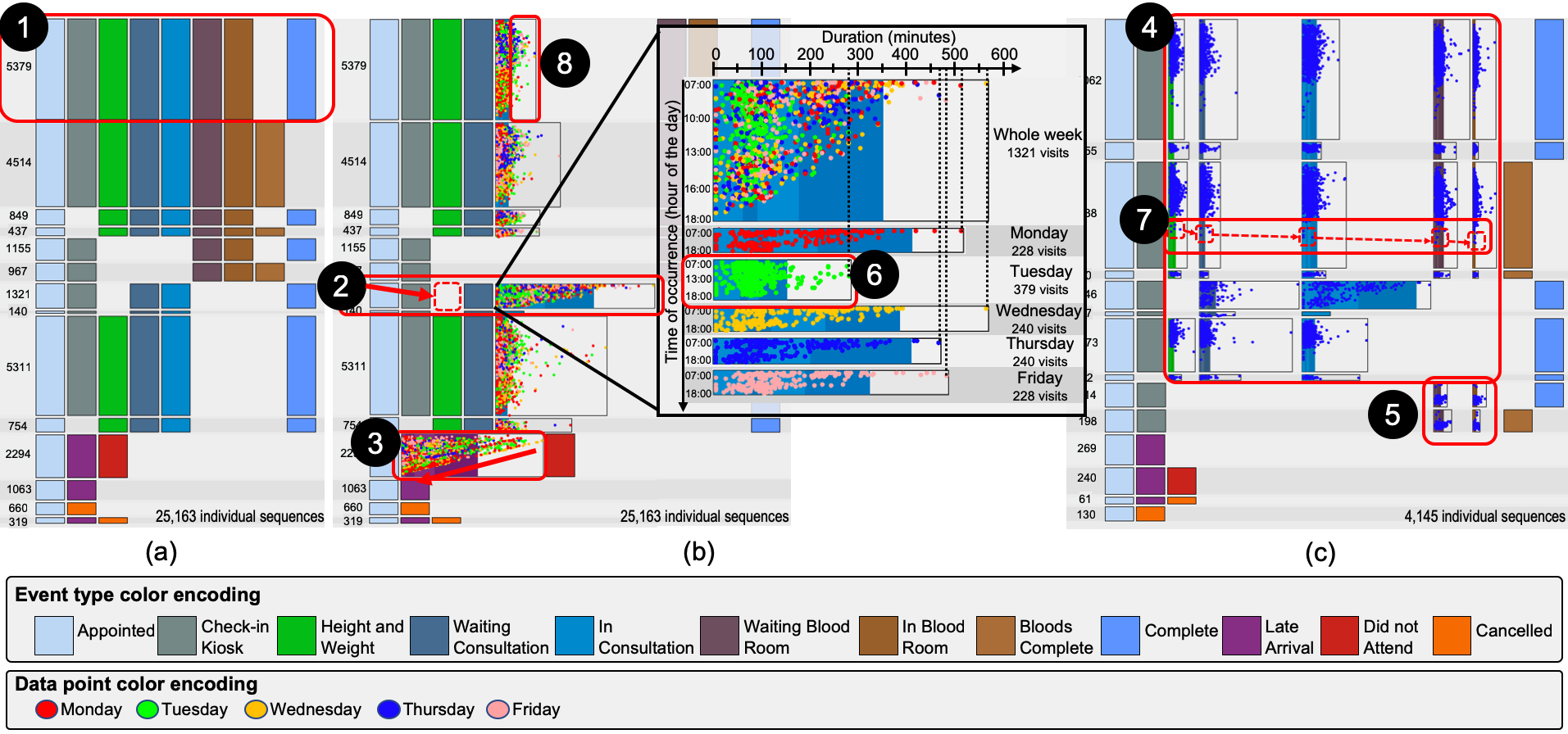}
   \caption{ Three different configurations for the Sequential and Time Patterns overview using the Rheumatology dataset. (a) Sequential and Time Patterns overview where all events are encoded as point events. (b) The events "In Consultation" and "Late Arrival" are expanded for analysis. The expanded events show the proposed event encoding for time attributes, which allow to identify trends and outliers with respect to duration and time of occurrence. For a selected sequence, the event "In Consultation" is broken down into days of the week for further exploration. (c) The Sequential and Time patterns overview is filtered to show only the events happening on Thursdays. The highlighted numbers represent findings which are explained in the taxonomy of findings proposed.}
   \label{fig:patterns}
}

\abstract{Event data is present in a variety of domains such as electronic health records, daily living activities and web clickstream records. Current visualization methods to explore event data focus on discovering sequential patterns but present limitations when studying time attributes in event sequences. Time attributes are especially important when studying waiting times or lengths of visit in patient flow analysis. We propose a visual analytics methodology that allows the identification of trends and outliers in respect of duration and time of occurrence in event sequences. The proposed method presents event data using a single Sequential and Time Patterns overview. User-driven alignment by multiple events, sorting by sequence similarity and a novel visual encoding of events allows the comparison of time trends across and within sequences. The proposed visualization allows the derivation of findings  that otherwise could not be obtained using traditional visualizations. The proposed  methodology has been applied to a real-world dataset provided by Sheffield Teaching Hospitals NHS Foundation Trust, for which four classes of conclusions were derived.} 

\CCScatlist{
   \CCScatTwelve{Hu\-man-cen\-tered com\-put\-ing}{Visu\-al\-iza\-tion}{Visu\-al\-iza\-tion techniques}{};
  \CCScatTwelve{Hu\-man-cen\-tered com\-put\-ing}{Visu\-al an\-a\-lyt\-ics}{}{};
  \CCScatTwelve{Ap\-plied com\-put\-ing}{Health care in\-for\-ma\-tion sys\-tems}{}{}
}

\begin{document}


\firstsection{Introduction}

\maketitle

Event logs are routinely recorded in a variety of domains such as electronic health records, daily living activities and web clickstream records. The analysis of event data can provide valuable insights into processes and the behaviour of individuals.

Existing visualization techniques focus on the visual encoding of sequential patterns, but present limitations when representing time attributes. Firstly, the time of occurrence is not explicitly encoded in the overview of sequences. Secondly, the duration of events (if visualized) is encoded using the average value. Exclusively representing average duration means that individuals with unusual durations (\textit{outliers}) are easily missed and, therefore, overlooked. The exact time at which individual events occur might not be of interest alone, but when combined with other variables, trends in time can be obtained to provide deeper insights. For example, visualizing time attributes is essential in the context of health care delivery. Waiting times and lengths of stay are key performance indicators required to optimize the use of resources whilst maximizing the quality of patient care.

We propose a generic methodology which is able to integrate the extraction and interactive visualization of sequential and time patterns from temporal event data. The identification of \textit{time patterns} is facilitated by a novel visual encoding capable of representing duration and time of occurrence \textit{at event level}. \textit{At sequence level}, a set of the most relevant unique sequences is obtained first, then sorted by sequence similarity, and finally visually aligned to obtain \textit{sequential patterns}.

\textcolor{black}{The proposed methodology allows users to interact with the data and focus their analysis on common patterns or on anomalous situations that might be of interest. We have applied the methodology to a real-world dataset in the healthcare domain, and present a taxonomy of findings that can be generalized to other application domains.} The contributions of the present work are:

\begin{itemize}[noitemsep]
  \item \textcolor{black}{\textbf{Visual encoding of time attributes at event level:} a novel visual encoding which aggregates the duration and time of occurrence of events.}
  \item \textcolor{black}{\textbf{Integration of time and sequential patterns into a single overview:} our approach allows the analysis of patterns that combine time attributes and the sequential order of event sequences.} 
\end{itemize}

\section{Related Work}
\subsection{\textcolor{black}{Visual analytics of Temporal Event Sequences}}
A variety of visual analytic techniques for event data have been proposed. Typically, an overview of the main sequential patterns is provided as a tree-like view, encoding sequences by frequency and average duration \cite{monroe2013temporal,wongsuphasawat2012exploring}. Other existing views include: icicle plots \cite{monroe2013temporal,liu2017coreflow}, sankey diagrams \cite{perer2014frequence,wongsuphasawat2012exploring}, transition matrices \cite{zhao2015matrixwave}, state transitions graphs \cite{vrotsou2009activitree,lam2007session}, and list of glyphs \cite{chen2018sequence}. Various techniques have addressed the identification of sequential patterns either by using frequent sequence mining (FSM) \cite{liu2017patterns,perer2014frequence,kwon2016peekquence} or clustering methods \cite{wongsuphasawat2012exploring,chen2018sequence,guo2018eventthread}.

The overview of sequential patterns can be manipulated using a wide range of operations which are application-dependent. 

In general, these operations can be grouped into three categories: transformations, queries and alignment. Transformations allow the simplification of the data overview \cite{monroe2013temporal}, for example either by merging similar event types into a single one or by removing records that are not of interest. Queries allow the filtering of data being visualized according to a set of events or temporal constraints of variable complexity. Existing strategies include: visual queries \cite{monroe2012exploring}, regular expressions \cite{cappers2018exploring} or milestone events \cite{liu2017coreflow,gotz2014decisionflow}. 
Alignment of event sequences by a selected event, target the exploration of the subset of events happening right before and after the alignment point. Typically, only one event can be selected as alignment point \cite{monroe2013temporal,chen2018sequence, wongsuphasawat2012exploring}, although more recently Eventpad \cite{cappers2018exploring} implemented a Multiple Sequence Alignment (MSA) strategy in which the user is able to modify the gap cost in the algorithm proposed by Bose et al.~\cite{bose2010trace}. 

\subsection{\textcolor{black}{Overview of time attributes}}
Existing visual analytics methods mainly focus \textcolor{black}{on} visualizing the sequential order of the events \cite{monroe2013temporal,perer2014frequence,wongsuphasawat2012exploring}. Currently, the visual encoding of time attributes is limited. Time of occurrence is always implicit in the sequential ordering of the sequences. However, no explicit time attributes (e.g. 3pm, Monday, May, 2019) are fully visually encoded in the overview. 

Generally, a separate secondary view is required to review the time of occurrence for a selected record. ActiviTree \cite{vrotsou2009activitree} visualizes the distribution of sequences across the time of the day, using a secondary view. In LifeLines2 \cite{wang2009temporal}, the distribution of the frequency of selected records through time is visualized using a histogram (i.e. the number of occasions a particular event happens on a specific date). Events within a time range before and after an alignment point can \textcolor{black}{also be} analyzed. However, this method does not aggregate sequences, and frequency distribution is only shown for the selected event.  

TimeSpan \cite{loorak2016timespan} uses stacked bar charts to indicate the duration of events related to a stroke treatment process, a line chart is used to study trends in duration through monthly intervals. However, they assume that sequences do not vary in the ordering of events, meaning sequential patterns are not included. 

In methods where the duration of events is visually encoded, the width of an event is scaled proportionally to the average duration.
This approach ignores the distribution of the duration and the presence of outliers. Duration outliers can be defined as observations with a duration which appears to be inconsistent with the remainder of the data \cite{Barnett1994}. 
Previous literature \cite{vrotsou2009activitree} indicates the importance of identifying infrequent sequences as outliers, but no emphasis has been made to pinpoint duration outliers. 

Some techniques offer the possibility of filtering events by their duration. Eventflow \cite{monroe2013temporal} allows querying event sequences after specifying a time window (e.g. displaying only events with a duration above thirty minutes), while Eventpad \cite{cappers2018exploring} provides a histogram to inspect event attributes separately. However, this information is not encoded in the sequential patterns overview. 
\section{Time attributes at event level}

Event data refers to a set of time-stamped events, where an \textit{event} can be expressed as what happened (\textit{event type}), when (\textit{time of occurrence}), for how long (\textit{duration}), and who was involved (\textit{identifier}). For each \textit{identifier}, an ordered list of events can be obtained (\textit{event sequence}); and for the complete set of event sequences in a dataset, the \textit{unique sequences} can be extracted by grouping those event sequences sharing the same ordered set of events exactly.

We propose a novel visual encoding to represent time attributes \textcolor{black}{at event level}, allowing the identification of trends and outliers with regard to duration and time of occurrence. The proposed approach also allows exploring time of occurrence at multiple time scales, and several levels of detail for the duration of events. 

\subsection{Visualizing duration and time of occurrence}

\label{sub:coreVisualization}

\begin{figure*}[h]
\centering
\subfloat[]{\includegraphics[height=4cm]{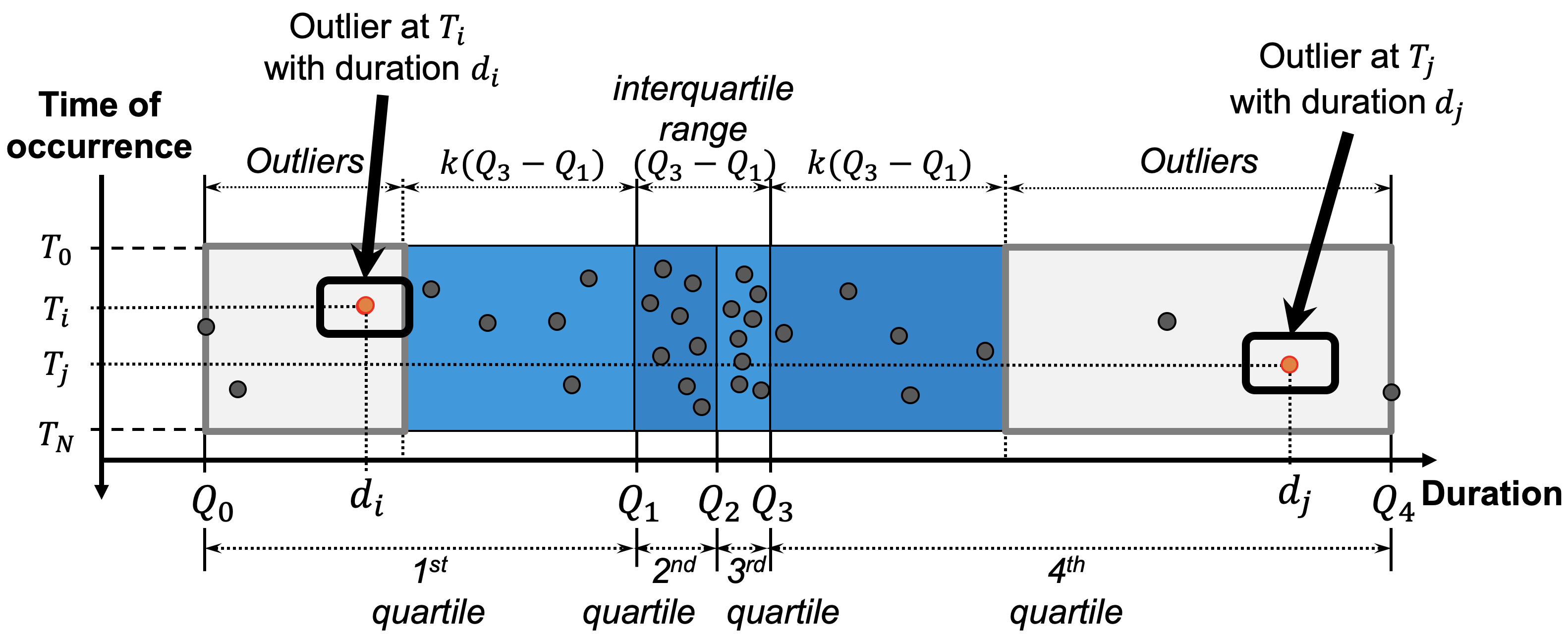}
\label{subfig:encodingEventType}}
\hfil
\subfloat[]{\includegraphics[height=5.2cm]{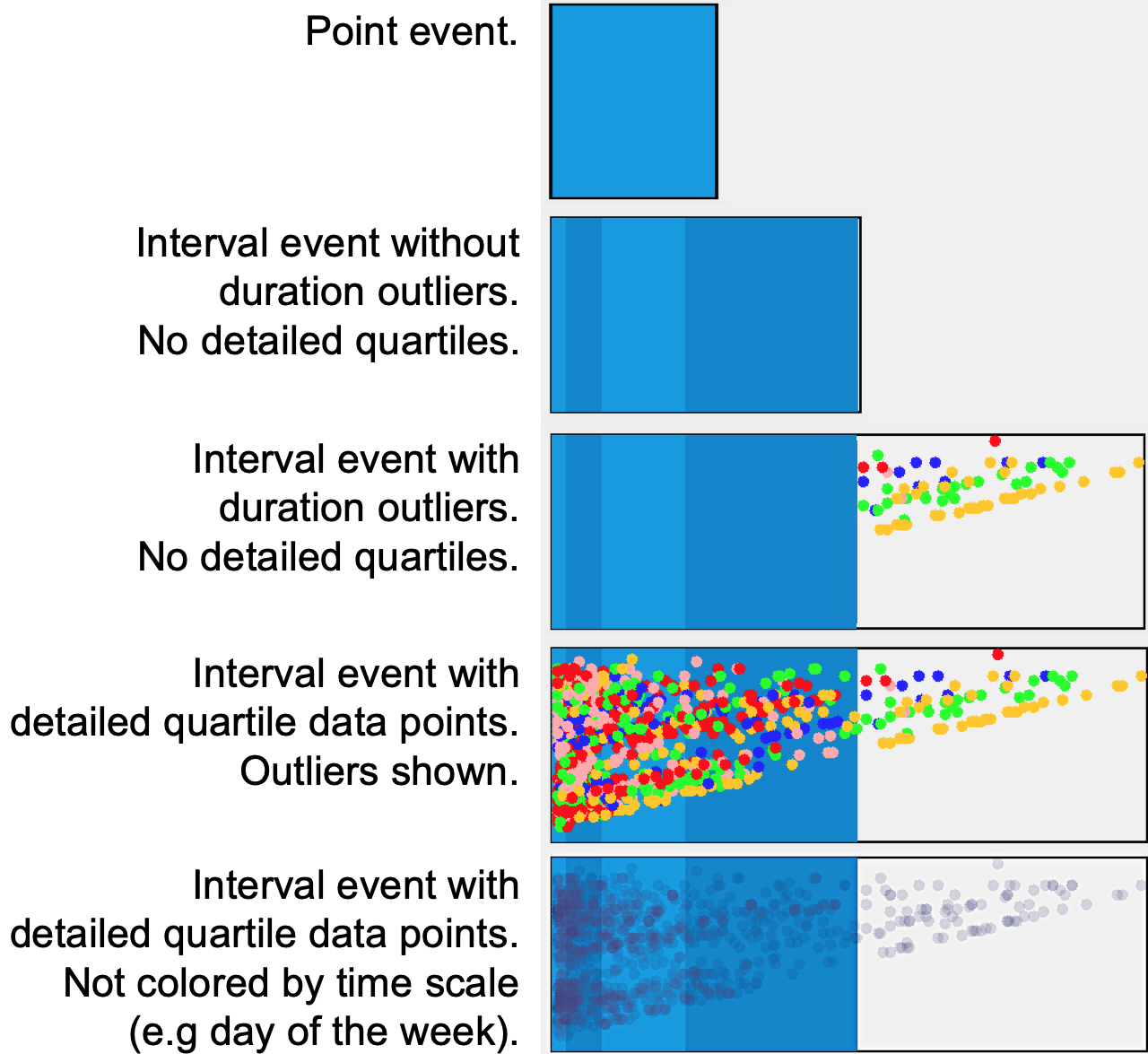}
\label{subfig:eventLevels}}
\caption{\protect \subref{subfig:encodingEventType} \textcolor{black}{Proposed visual encoding of an event box which is used to aggregate events. Data points represent individual event occurrences. Quartiles are delimited by duration ranges, where $Q_0$ is the minimum duration and $Q_4$ the maximum. Data points are located in the horizontal axis according to duration and in the vertical axis according to time of occurrence.} \protect \subref{subfig:eventLevels} \textcolor{black}{Example of an event box at five different levels of detail. Hiding or showing quartile and outlier data points, and changing the colour of the data points, result in different levels of detail.}}
\label{fig:timeAttribute}
\end{figure*}

The proposed visual encoding of \textcolor{black}{events} is inspired in boxplots \cite{tukey1977boxplot}. For univariate data, boxplots have proved excellent to understand a whole sample of observations. They present multiple benefits: highlight of outliers, resistance to extreme values, representation of variability and skewness of the sample. 

\textcolor{black}{An \textit{event box} aggregates events of a particular type (see \autoref{subfig:encodingEventType}), its \textit{height} is proportional to the number of records, and its \textit{width} proportional to the maximum variation ($Q_4-Q_0$) in duration. An event box is composed by \textit{quartiles} and \textit{data points}.}

\textcolor{black}{Quartiles are represented as colored sub boxes} delimited by a duration value $Q_i$. In this case, $Q_0$ is the minimum duration and $Q_4$ the maximum duration. $Q_1$,$Q_2$ and $Q_3$ correspond to the 25\textsuperscript{th}, 50\textsuperscript{th} and 75\textsuperscript{th} percentile respectively. Each data point in the visualization represents an individual event occurrence. Depending on their duration, data points can be either considered as \textit{outliers} or \textit{quartile points}. 

Using Tukey's \cite{frigge1989some} definition, outliers are identified as data points outside the range:
$$[Q_1-k(Q_3-Q_1),Q_3+k(Q_3-Q_1)],$$ 
where $(Q_3-Q_1)$ is the interquartile range and traditionally $k=1.5$. \textcolor{black}{Opposite to outliers, quartile points are visually identified as those within the colored sub boxes.} 

Data points are placed along the \textit{horizontal} axis according to their duration, and placed in the \textit{vertical} axis according to their time of occurrence. The scale of the \textit{vertical} axis is given by the range $T_0$ to $T_N$, where $T_0$ is the minimum and $T_N$ represents the maximum time of occurrence being visualized. This arrangement of data points allows the study of the distribution of duration as well as time of occurrence over a given period of time.

\subsection{Time of occurrence at multiple time scales}\label{sub:multipleTimeScales}
The proposed visual encoding can be customized to show time at multiple scales. The units for the time of occurrence (vertical axis) shown in \autoref{subfig:encodingEventType} can be adjusted to show \textcolor{black}{the time interval} $[T_0,T_N]$, as either hours of the day, days of the week, months, or even years.

The use of color adds a further dimension to the visualization. For example the vertical axis can represent the hour of the day, with a range starting from $T_0=8am$ to $T_N=5pm$ at 1-hour intervals; whilst data points can be simultaneously colored according to the day of the week (i.e. data points in red occur on Mondays and blue on Fridays). Combinations of multiple time scales between the vertical axis and the data point color are possible.

\subsection{Levels of detail in the duration of events}\label{sub:levelsOfDetail}
In terms of duration, events are typically categorized into either point or interval events \cite{monroe2012exploring}. 

\textcolor{black}{The proposed event box visualization can be customized to show, hide or modify the detail of data points. This allows to produce the following levels of detail (see \autoref{subfig:eventLevels}):}
\begin{itemize}[noitemsep]
    \item\textit{Point event}: event details are fully \textit{collapsed} to reduce visual clutter.
    \item\textit{Interval event without duration outliers}: duration outliers (if any) are hidden.
    \item\textit{Interval event with duration outliers}: outliers (if any) are shown.
    \item\textit{Interval event, detailed quartiles}: data points inside colored sub boxes are shown.
    \item\textit{Interval event, no detailed quartiles}: data points inside colored sub boxes are hidden.
    \item\textit{Interval event not colored by time scale}: Data points are colored using transparency, so that the volume of the points is observed.
\end{itemize}
The above will be used to help reducing the visual clutter during the analysis of data that will be discussed in the following section. Transforming events into point or interval events allow to focus the analysis of duration in selected events. Sometimes the user might be interested in the occurrence of an event without being interested in its duration. A point event will provide context without having to remove it from display.

\section{Time attributes across sequences}
Our methodology is able to seamlessly integrate the extraction and interactive visualization of sequential and time patterns from temporal event data. 
\subsection{Sequential and Time Patterns overview}
\textcolor{black}{Sequential and time patterns are integrated in a single overview, which is built in four steps: 1) the unique sequences that explain most of the variability in the original dataset are selected, resulting in the removal of unique sequences with relatively low frequency; 2) the selected unique sequences are sorted by similarity using complete-link agglomerative hierarchical clustering \cite{hartigan1975clustering} and the Levenshtein edit distance \cite{levenshtein1966binary}; 3) sequences are aligned by multiple events as selected by the user; 4) the events in each unique sequence are represented using the visual encoding in \autoref{subfig:encodingEventType}.
This type of overview allows comparison of time attributes across sequences. 
\autoref{fig:patterns} presents three configurations of this overview for the same dataset using different visualization settings and filters.}

\subsection{Multiple alignment and user interaction}
\textcolor{black}{The present methodology allows users to interact with the data and focus their analysis on common patterns or on anomalous situations that might be of interest. A variety of interaction mechanisms are offered to manipulate and explore the overview of Sequential and Time Patterns:}
\begin{itemize}[noitemsep]
\item \textit{User-driven multiple alignment:} When new alignment events are selected, sequences are re-sorted by the similarity to the alignment events and by similarity within sequences.
\item \textit{Manipulation of event encoding to reduce visual clutter:} User can change the level of detail of selected events as per \autoref{sub:levelsOfDetail}.
\item \textit{Filter by date:} Records can be filtered by date range or specific days of the week. The data points described in \autoref{subfig:encodingEventType} can be colored according to the day of the week.
\item \textit{Breaking down a unique sequence:} A unique sequence can be subdivided according to the day of the week or other criteria.
\end{itemize}

\section{Case study and Taxonomy of findings}
To demonstrate the capabilities of our methodology, we selected a dataset in the healthcare domain as our case study.  From the analysis of this dataset, four main classes of findings were identified. We propose a taxonomy of findings which can be generalized to other application domains.

\subsection{Case study}
The understanding of patient flow is an area where good data analysis is critical \cite{bardsley2019untapped}. Waiting times, lengths of stay and clinical pathways are key aspects to the study of patient flow. 

In this work, we used one year of real-world patient flow data from a Rheumatology outpatient clinic (Sheffield Teaching Hospitals - NHS Foundation Trust, Sheffield, United Kingdom). On average, the Department has an approximate annual load of 9000 patients and 25000 appointments. 
Patient visits at this clinic are routinely tracked using an in-house workflow tracking system\textcolor{black}{, where the clinical staff (e.g nurses, receptionist, consultants) input the current state of a patient according to the service being provided.} The produced event logs are used by the hospital to obtain basic statistics about the quality of care being delivered, particularly focusing on the study of waiting times and lengths of visit. 

Our analyses offered the possibility of delving into the raw event logs to extract key insights about patient flow within the clinic (see specific dataset details in \autoref{tab:dataSpec}).  These are being used to gain a better understanding of how the department operates and to suggest strategies about how to optimize the delivery of care. 
\begin{table}[h]
\caption {Characteristics of the Rheumatology dataset.} \label{tab:dataSpec} 
\centering
\begin{tabular}{ lcccc }
\toprule
\bfseries No. event & \bfseries Total no.     & \bfseries No. unique    & \bfseries Time      \\
\bfseries  types     & \bfseries sequences     & \bfseries sequences     & \bfseries period    \\ 
\midrule
18        & 26,455        & 161           & 1 year    \\
\bottomrule
\end{tabular}
\end{table}

\subsection{Taxonomy of findings and examples}
The generic taxonomy of findings is presented in this section.
This taxonomy comprises four classes of findings.  Examples for each class are provided using the Rheumatology dataset.

\textbf{Class I - Discovering main sequential patterns}: The user can visually identify the main and most representative sequential patterns in the Sequential and Time Patterns overview. Examples of this class are:
\begin{itemize}
    \item The most frequent unique sequence is the clinical pathway Appointed $\rightarrow$ Check-in Kiosk $\rightarrow$ Height and Weight $\rightarrow$ Waiting Consultation $\rightarrow$ In Consultation $\rightarrow$ Waiting Blood Room $\rightarrow$ In Blood Room $\rightarrow$ Complete (Finding 1 in \autoref{fig:patterns}a). 
\end{itemize}

\textbf{Class II - Impact of an event in the duration of subsequent events}: The proposed sorting and alignment methodology allows detecting differences and relations among sequences. The variation in the duration of specific events across sequences could be related to the presence or absence of certain events. For example, the sequences "ABCDE" and "BCDFG", share the subsequence "BCD". This class of finding suggests that if the duration of "BCD" varies between the two sequences, that variation might be \textcolor{black}{associated} to the occurrence of the event "A". This finding should require further investigation to explain the situation.

Examples of this class are: 
\begin{itemize}
    \item The time of the event In Consultation is considerably longer when the event Height and Weight does not occur in that sequence (Finding 2 in \autoref{fig:patterns}b). When we interviewed the clinical staff, they indicated that those patients, \textcolor{black}{without Height and Weight,} visit the clinic to undergo longer day-case procedures rather than conventional consultations. \textcolor{black}{A day-case procedure will likely include a consultation plus extra activities such as intravenous infusion of drugs, X rays and other imaging tests; hence the long duration of this event}.
\end{itemize}

\textbf{Class III - Trends in duration with regard to time of occurrence}: The proposed visualization allows to identify trends of duration through time. For example, if the duration increases as the time elapses or if high durations occur only in the early hours or a specific day.  Examples of this class are:
\begin{itemize}
    \item In Finding 3 in \autoref{fig:patterns}b, the duration of the late arrival event decreases as the day goes by; meaning that the amount of minutes that patients are late are higher in the morning than towards the end of the day. The highest durations (4th quartile and outliers) are concentrated in the morning (between 9:00 and 12:00). This requires a further investigation of morning appointments, revising the reasons for morning late arrivals.
\end{itemize}

\textbf{Class IV - Temporal distribution of time attributes}: The proposed visualization helps identifying what is the "normal" duration of an event, as well as identifying the distribution of points through time of occurrence. Examples of this class are: 
\begin{itemize}[noitemsep]
    \item Distribution of time of occurrence: On Thursdays, the visits to the clinic feature a higher concentration before noon (Finding 4 in \autoref{fig:patterns}c). However, the visits for which the purpose is exclusively a blood test occur in the afternoon (Finding 5 in \autoref{fig:patterns}c).
    \item Distribution of duration: On Tuesdays, consultation times are significantly shorter, which suggests that the clinic running on that day might be dealing with less complex pathologies (Finding 6 in \autoref{fig:patterns}b).  
    \item Unusual times of occurrence: On Thursdays, the majority of patients visit the clinic in the mornings. Nevertheless, a reduced number of patients are seen in the afternoons (Finding 7 in \autoref{fig:patterns}c).  Investigating that cohort of patients would be of interest.
    \item Unusual duration: The present method allows for the identification of patients that stay in an event for an unusual amount of time. \textcolor{black}{Patients that have stayed in an event for an unusual time are represented as outlier data points in the proposed visualization, these are visually identified as points outside the colored sub boxes (e.g. see highlighted outliers in \autoref{fig:patterns}, Finding 8)}.
\end{itemize}

\section{Discussion and Conclusion}
This paper proposes a generic methodology to analyze time attributes in event data, at event level and across event sequences. 
A Sequential and Time Patterns overview visualization is built after extracting the most relevant unique sequences. To simplify the exploration of temporal relations among sequences, unique sequences are sorted by similarity and aligned by multiple events selected by the user.

To facilitate the understanding of how events relate across sequences and within sequences, a novel visual encoding of duration and time of occurrence is embedded in the Sequential and Time patters overview. This encoding enables users to gain valuable insights in to how events distribute with regard to time attributes.  To make this possible, users are able to interact with the data at multiple levels of detail and time scales using a single overview. 

Based on the presented case study in the healthcare domain, a taxonomy of findings generalizable to other applications has been proposed. The four classes of conclusions presented have already opened the possibility of identifying patterns of interest that should be further investigated to support hospital service improvement.

\textcolor{black}{Depending on the application domain, the methodology could present limitations in terms of scalability. Real-world datasets can contain hundreds or thousands of unique sequences, future work is necessary to create summaries of unique sequences at different levels of detail. Other scalability issues could be present as the ranges of duration $Q_4-Q_0$ and time of occurrence $T_N-T_0$ become larger, or as the volume of the data points in an event box increase. To solve these potential limitations: time windows, zoom and scaling functions could be used.}


\bibliographystyle{abbrv}

\bibliography{mybibliography}
\end{document}